\documentclass[a4paper,11pt]{article}
\pdfoutput=1 
\usepackage{jheppub} 
\usepackage[T1]{fontenc} 
\usepackage{multirow}
\usepackage{color}
\usepackage{ulem}

\usepackage{mathrsfs}

\title{\protect\boldmath Measurement of $\Lambda$ transverse polarization in $e^{+}e^{-}$ collisions at $\sqrt{s}= 3.68 - 3.71$ GeV}

\collaborationImg{\includegraphics[width=0.15\textwidth]{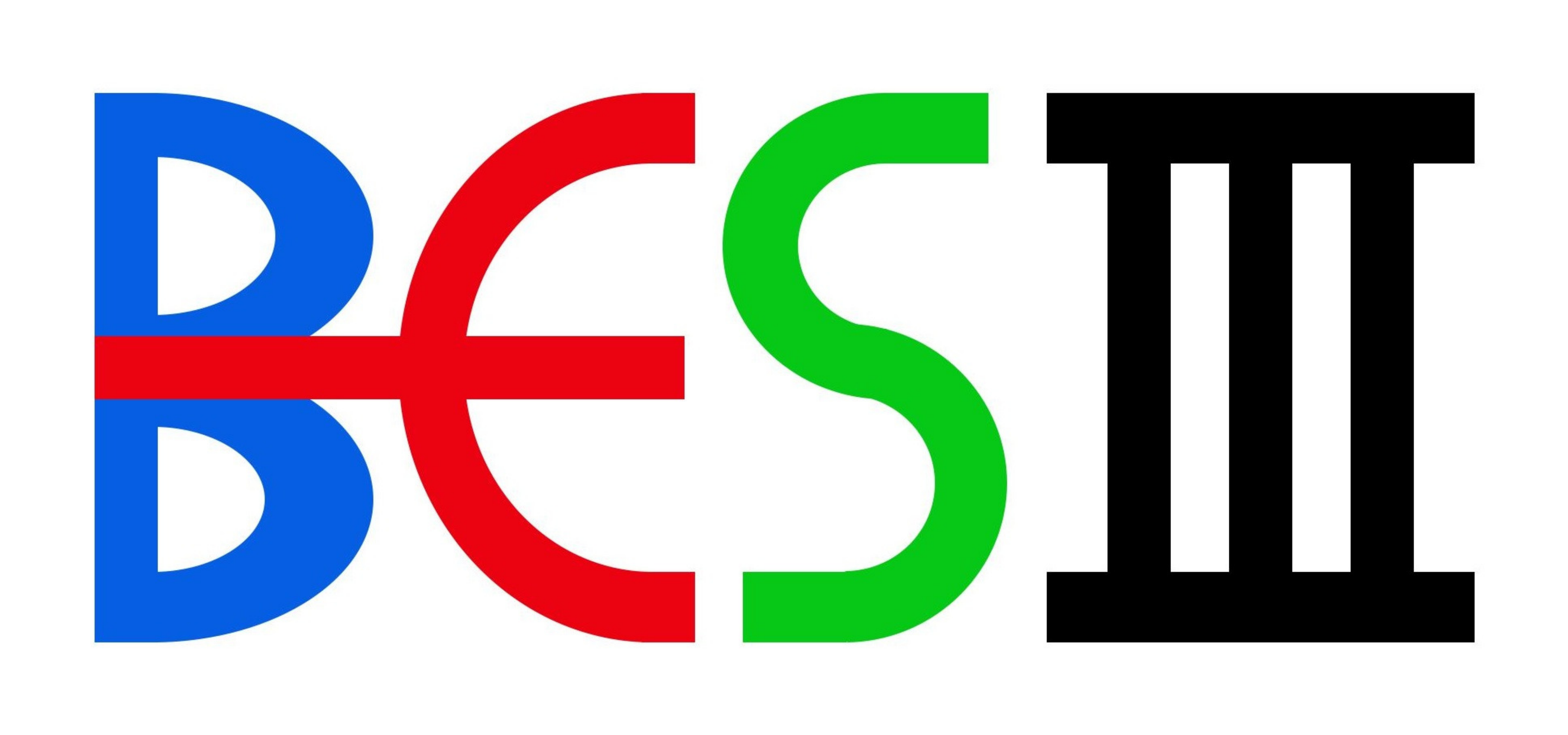}}
\collaboration{The BESIII Collaboration}

\author{
M.~Ablikim$^{1}$, M.~N.~Achasov$^{13,b}$, P.~Adlarson$^{73}$, R.~Aliberti$^{34}$, A.~Amoroso$^{72A,72C}$, M.~R.~An$^{38}$, Q.~An$^{69,56}$, Y.~Bai$^{55}$, O.~Bakina$^{35}$, R.~Baldini Ferroli$^{28A}$
% ...
\\
\vspace{0.2cm}
(BESIII Collaboration)\\
\vspace{0.2cm} {\it
$^{1}$ Institute of High Energy Physics, Beijing 100049, People's Republic of China\\
% ...
$^{13}$ G.I. Budker Institute of Nuclear Physics SB RAS (BINP), Novosibirsk 630090, Russia\\
% ...
$^{28}$ INFN Laboratori Nazionali di Frascati , (A)INFN Laboratori Nazionali di Frascati, I-00044, Frascati, Italy; (B)INFN Sezione di  Perugia, I-06100, Perugia, Italy; (C)University of Perugia, I-06100, Perugia, Italy\\
% ...
$^{34}$ Johannes Gutenberg University of Mainz, Johann-Joachim-Becher-Weg 45, D-55099 Mainz, Germany\\
$^{35}$ Joint Institute for Nuclear Research, 141980 Dubna, Moscow region, Russia\\
% ...
$^{38}$ Liaoning Normal University, Dalian 116029, People's Republic of China\\
% ...
$^{55}$ Southeast University, Nanjing 211100, People's Republic of China\\
$^{56}$ State Key Laboratory of Particle Detection and Electronics, Beijing 100049, Hefei 230026, People's Republic of China\\
% ...
$^{69}$ University of Science and Technology of China, Hefei 230026, People's Republic of China\\
% ...
$^{72}$ University of Turin and INFN, (A)University of Turin, I-10125, Turin, Italy; (B)University of Eastern Piedmont, I-15121, Alessandria, Italy; (C)INFN, I-10125, Turin, Italy\\
$^{73}$ Uppsala University, Box 516, SE-75120 Uppsala, Sweden\\
% ...
$^{b}$ Also at the Novosibirsk State University, Novosibirsk, 630090, Russia\\
% ...
}
}
\emailAdd{besiii-publications@ihep.ac.cn}

\begin{document} 

\abstract{
With data samples collected with the BESIII detector at seven energy points at $\sqrt{s}= 3.68 - 3.71$ GeV, corresponding to an integrated luminosity of 333 pb$^{-1}$, we present a study of the $\Lambda$ transverse polarization in the $e^+e^-\to\Lambda\bar\Lambda$ reaction. The significance of polarization by combining the seven energy points is found to be 2.6$\sigma$ including the
systematic uncertainty, which implies a non-zero phase between the transition amplitudes of the $\Lambda\bar\Lambda$ helicity states.
The modulus ratio and the relative phase of EM-$psionic$ form factors combined with all energy points are measured to be 
$R^{\Psi} =$ 0.71$^{+0.10}_{-0.10}$ $\pm$ 0.03 and 
$\Delta\Phi^{\Psi}$ = 23$^{+8.8}_{-8.0}$ $\pm$ 1.6$^\circ$,
where the first uncertainties are statistical and the second systematic.}

\maketitle
\flushbottom
\section{Introduction}
\label{sec:intro}

The understanding of the structure of baryons is a very important issue in contemporary physics~\cite{Brodsky:1974vy, Geng:2008mf, Green:2014xba, Wang:2022zyc,
Wang:2022bzl, Liu:2023xhg}.
In the context of the Quantum Chromodynamics (QCD), it is particularly interesting to measure the electromagnetic form factors (EMFFs) of nucleons and hyperons, which are expected to reveal the aspects of the QCD description of the hyperon structure.
In the 1960s, Cabibbo {\it et al.}~\cite{Cabibbo:1961sz} first proposed that time-like EMFFs could be studied on $e^+e^-$  experiments by measuring the {baryon} pair
production cross sections. Among them, the proton is a stable particle, and can be available as a target to study its space-like EMFFs by means of scattering experiments.
This case is different from the unstable hyperons with finite lifetime which cannot be used in such scattering experiments. 
The advantage is that their weak parity-violating decay gives straightforward access to the polarization. 
 The time-like form factors are related to more intuitive quantities such as charge and magnetization densities by dispersion relations~\cite{Belushkin:2006qa, Yan:2023yff}. 
The production of spin-1/2 baryon-anti-baryon pairs from $e^+e^-$ collisions is described by two independent parameters, the electric form factor $G_{E}$ and the magnetic form factor $G_{M}$~\cite{Huang:2021xte, Qian:2021neg}.

They are both analytic functions of the four-momentum transfer squared $q^2$. 
In the time-like region, starting from the threshold, corresponding to the squared mass of the lightest hadronic state that can couple to the intermediate virtual photon, the EMFFs are complex. In particular, the complex value of their ratio implies a polarization effect in the final state baryons even when the initial state leptons are unpolarized. 
This provides a handle to understand the intrinsic structure of hyperons better.

Up to now, experimental data on hyperon EMFFs are very limited. The first 
determination of $\Lambda$ EMFFs was reported by the BABAR collaboration~\cite{Aubert:2007uf} using the initial state radiation (ISR) method for the $e^+e^- \to \Lambda\bar\Lambda$ process. It measured the effective form factor, which is proportional to the total cross section assuming one-photon exchange. The cross section and EMFFs of some baryon pairs ($p$, $\Lambda$, $\Sigma^0$, $\Xi^-$, and $\Omega^-$) were determined by the CLEO collaboration~\cite{Dobbs:2014ifa,Dobbs:2017}. Their conclusions regarding EMFFs and di-quark correlations~\cite{Jafee:2005,Jafee:2003} rely on the assumption that one-photon exchange dominates the production process and that decaying charmonia contributions are negligible. 
The BESIII collaboration has also measured cross sections of some baryon pairs ($\Lambda$, $\Sigma^0$, $\Xi^-$, $\Sigma^\pm$, $\Xi^0$ and $\Omega^-$) near the production threshold~\cite{BESIII:2017hyw, BESIII:2020uqk, BESIII:2020ktn, BESIII:2021aer, BESIII:2021rkn, BESIII:2022kzc}
and above the open charm threshold~\cite{BESIII:2019cuv, BESIII:2023rse, BESIII:2021ccp}, 
while the experimental investigations of the relative phase between $G_E$ and $G_M$ are still limited.

At the resonances of vector charmonia, the spin formalism~\cite{Faldt:2017kgy} is valid. In these cases, the amplitudes no longer represent EMFFs but instead the so-called EM-\textit{psionic} form factors, $G^{\Psi}_E$ and $G^{\Psi}_M$~\cite{BESIII:2021cvv}. And the polarization is determined by the relative difference of electric and magnetic form factors $\Delta\Phi^{\Psi} \equiv \Phi^{\Psi}_{E} - \Phi^{\Psi}_{M}$, {with $G_{E,M}=\left|G_{E,M}\right|e^{i\Phi_{E,M}}$}, which were neglected in previous studies~\cite{BESIII:2019cuv, Ablikim:2016iym, Ablikim:2016iym-01, Ablikim:2016iym-02, Ablikim:2016iym-03, BESIII:2020ktn, BESIII:2021aer, Wang:2018kdh, Wang:2021lfq,
BESIII:2022mfx}. Recently the $\Lambda$ polarization was observed and measured in the $e^+e^- \to \Lambda\bar\Lambda$ process by the BESIII collaboration in $J/\psi$, $\psi(3770)$ and off-resonance regions~\cite{Ablikim:2018zay, BESIII:2021cvv, BESIII:2022yprl, BESIII:2019nep}.
Subsequently, the $\Sigma^+$ hyperon polarization was observed by the BESIII collaboration in $e^+e^-\to J/\psi, \psi(3686)\to\Sigma^+\bar{\Sigma}^-$ processes~\cite{yanliang}. The results reveal not only  a non-zero relative psionic phase, but also that the phase changes sign at the $\psi(3686)$ mass with respect to the value measured at the $J/\psi$ resonance. In the $e^+e^-\to J/\psi,\psi(3686)\to\Xi^-\bar\Xi^+$ channel~\cite{BESIII:2021ypr,BESIII:2022_xi_psip, BESIII:2023lkg}, a non-zero polarization has also been observed for the $\Xi^{-}$ hyperon.
The energy points around 3.686 GeV are interesting in this regard since the production occurs through an interplay of one-photon exchange~\cite{BESIII:2019nep}, mixing with $\psi(3770)$ resonance~\cite{BESIII:2021cvv} and resonance dominating only~\cite{Ablikim:2018zay,BESIII:2022yprl}.
The large data samples corresponding to an integrated luminosity of 333 pb$^{-1}$, collected at $\sqrt{s} =$ 3.680, 3.683, 3.684, 3.685, 3.687, 3.691, and 3.710 GeV with the BESIII detector~\cite{Ablikim:2009aa} recording symmetric $e^+e^-$ collisions provided by the BEPCII storage ring~\cite{Yu:IPAC2016-TUYA01}, enable the study of this phenomenon, which we present in this article.

\section{BESIII detector and Monte Carlo simulation}
The BESIII detector~\cite{Ablikim:2009aa} records symmetric $e^+e^-$ collisions 
provided by the BEPCII storage ring~\cite{Yu:IPAC2016-TUYA01}
in the center-of-mass (CM) energy range from 2.0 to {4.95~GeV},
with a peak luminosity of $1 \times 10^{33}\;\text{cm}^{-2}\text{s}^{-1}$ 
achieved at $\sqrt{s} = 3.77\;\text{GeV}$. 
BESIII has collected large data samples in this energy region~\cite{Ablikim:2019hff}. The cylindrical core of the BESIII detector covers 93\% of the full solid angle and consists of a helium-based
 multilayer drift chamber~(MDC), a plastic scintillator time-of-flight
system~(TOF), and a CsI(Tl) electromagnetic calorimeter~(EMC),
which are all enclosed in a superconducting solenoidal magnet
providing a 1.0~T magnetic field.
{
The magnetic field was 0.9~T in 2012, which affects 100\% of the total $J/\psi$ data.}
The solenoid is supported by an
octagonal flux-return yoke with resistive plate counter muon
identification modules interleaved with steel. 
The charged-particle momentum resolution at $1~{\rm GeV}/c$ is
$0.5\%$, and the 
${\rm d}E/{\rm d}x$
resolution is $6\%$ for electrons
from Bhabha scattering. The EMC measures photon energies with a
resolution of $2.5\%$ ($5\%$) at $1$~GeV in the barrel (end cap)
region. The time resolution in the TOF barrel region is 68~ps, while
that in the end cap region is 110~ps. {The end cap TOF
system was upgraded in 2015 using multigap resistive plate chamber
technology, providing a time resolution of
60~ps,
which benefits 100\% of the data used in this analysis~\cite{etof1,etof2,etof3}.}

Simulated data samples produced with a {\sc geant4}-based~\cite{GEANT4} Monte Carlo (MC) package, which includes the geometric description of the BESIII detector~\cite{Huang:2022wuo} and the detector response, are used to determine detection efficiencies and to estimate backgrounds. The simulation models the beam energy spread and ISR in the $e^+e^-$ annihilation using the generator {\sc kkmc}~\cite{KKMC}.
The detection efficiency for $e^+e^-\to\Lambda\bar\Lambda$ is determined by MC simulations. A sample of 1,000,000 events is simulated with a uniform phase space (PHSP) distribution for each of 7 CM energy points from 3.68 GeV to 3.71 GeV.
The $\Lambda$ baryon and its subsequent decays are handled by the {\sc evtgen} program~\cite{evtgen2,EVTGEN} with PHSP model. 
The production process is simulated by the {\sc kkmc} generator that includes the beam energy spread and ISR\cite{continuum} in the $e^+e^-$ annihilation.

\section{Event selection}
The full reconstruction method is performed to proceed event selection with the decay processes $\Lambda\to p\pi^- $
and $\bar{\Lambda}\to \bar{p}\pi^+$.
There are four charged particles in the final state, {a} proton, {an} anti-proton {and} two charged pions from
$\Lambda\bar{\Lambda}$.  Thus, good candidates should satisfy the event selection criteria below.

Charged tracks are required to be reconstructed in the MDC within its angular coverage: $|\cos\theta|<0.93$, where $\theta$ is defined with respect to the $z$ axis, which is the symmetry axis of the MDC in the laboratory system. Events with at least two negative-charged tracks and two positive-charged tracks are kept for further analysis. 
{Tracks with momentum larger than $0.6~{\rm GeV}/c $ are considered as proton candidates, and others are assumed to be pion candidates.}
Events with at least one proton, one anti-proton, one $\pi^+$, and one $\pi^-$ are retained for further analysis.

To reconstruct $\Lambda(\bar\Lambda)$ candidates, a secondary vertex fit~\cite{XUM} is applied to all combinations of $p\pi^{-}(\bar{p}\pi^{+})$ within one event. The pair of $\Lambda$ and $\bar\Lambda$ candidates  with the minimum value
of $\sqrt{(M_{p\pi^{-}}-m_{\Lambda})^{2} + (M_{\bar{p}\pi^{+}}-m_{\Lambda})^{2}}$ is selected.
Here, $M_{p\pi^{-}(\bar{p}\pi^{+})}$ is the invariant mass of the $p\pi^{-}(\bar{p}\pi^{+})$ pair. 
To further suppress background from non-$\Lambda$ events, the decay length of $\Lambda$ candidate,  i.e., the distance between its production and decay positions, is required to be greater than zero. 

\begin{figure}[!htbp]
	\begin{center}
\includegraphics[width=0.6\textwidth]{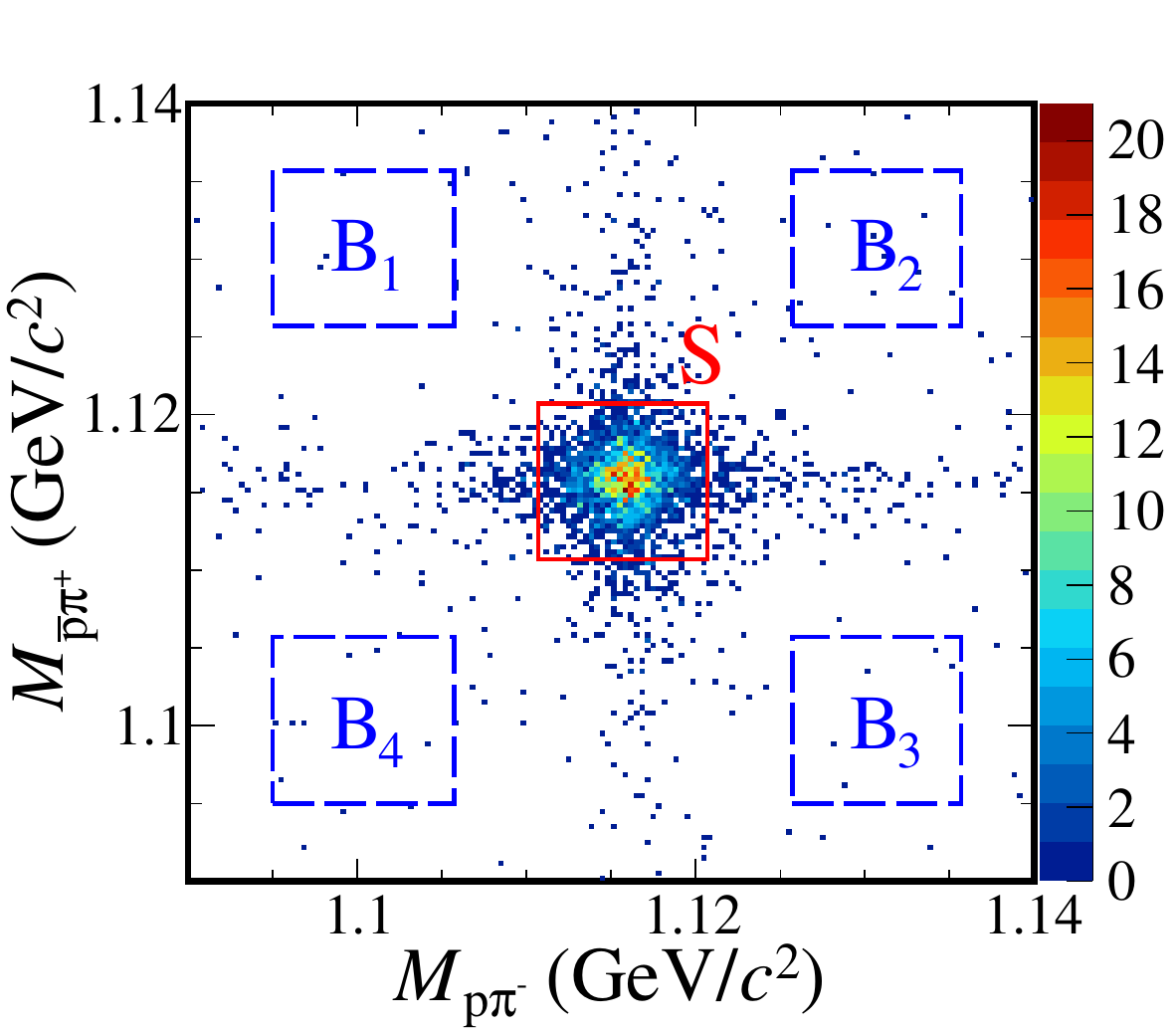}
	\end{center}
\caption{Two-dimensional distribution of $M_{\bar{p}\pi^{+}}$ versus $M_{p\pi^{-}}$ for all data samples, where the {red solid} box indicates the signal region, the {blue dash} boxes show the selected background regions.
}
\label{scatter_plot::llb}
\end{figure}

To further  suppress background contributions and improve the mass resolution, a four-constraint (4C) kinematic fit imposing energy-momentum conservation from the initial $e^+e^-$ to the final $\Lambda\bar\Lambda$ state is applied for all $\Lambda\bar\Lambda$ hypotheses after the $\Lambda(\bar\Lambda)$
reconstruction, with the requirement of $\chi^{2}_{\rm 4C} < 200$. Figure~\ref{scatter_plot::llb} shows the distribution of
$M_{\bar{p}\pi^{+}}$ versus $M_{p\pi^{-}}$ after performing the 4C kinematic fit. A clear accumulation of events around $m_{\Lambda}$ can be seen.

\section{Extraction of $\Lambda$ polarization}
\begin{figure}[!htbp]
	\begin{center}
\includegraphics[width=0.6\textwidth]{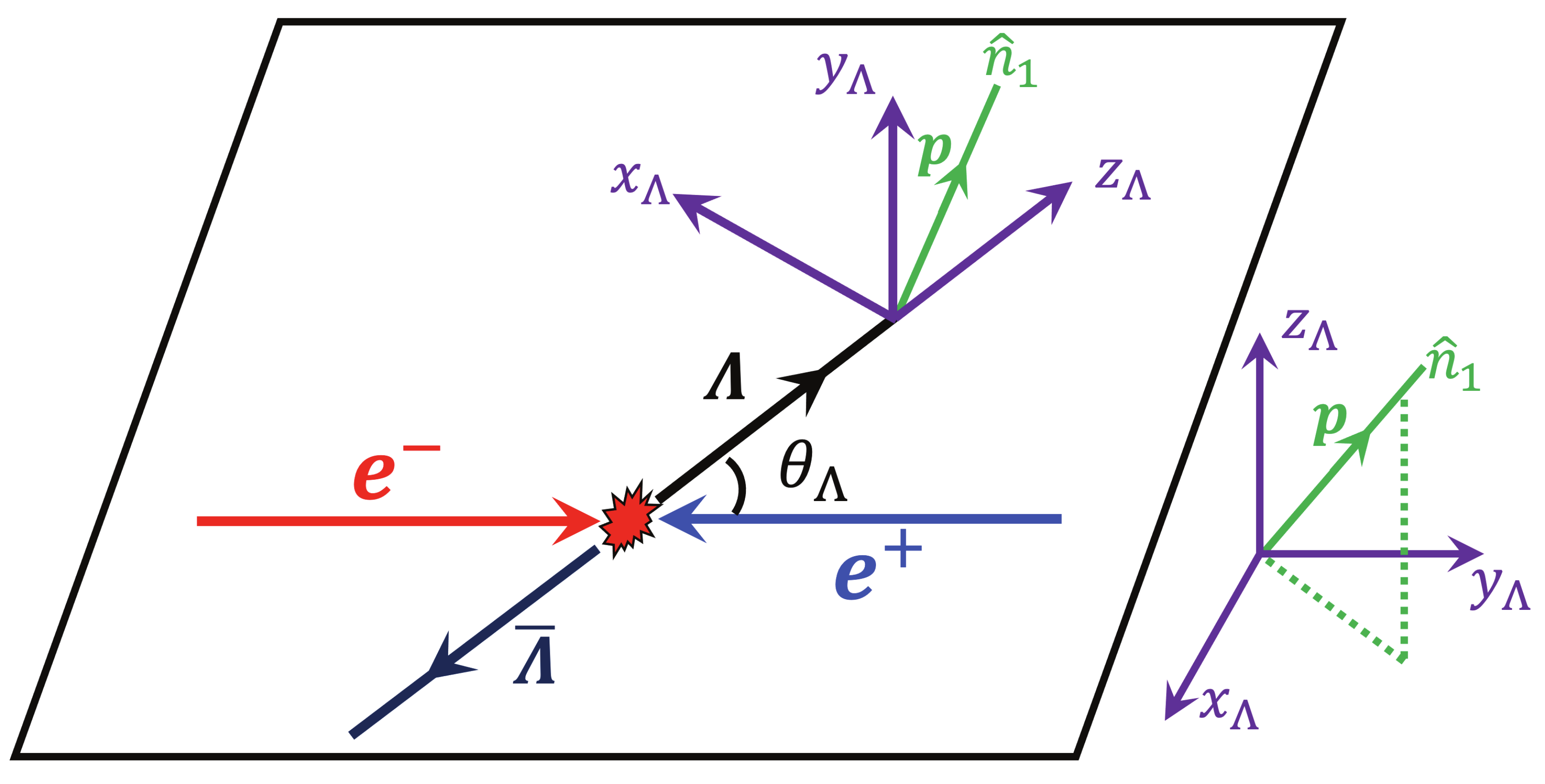}
	\end{center}
\caption{
Definition of the coordinate system used to describe the {$e^+e^-\to\gamma^*/\Psi\to\Lambda\bar\Lambda\to p\bar{p}\pi^{+}\pi^{-}$ reaction}. The $\Lambda$ particle is emitted along the $z_{\Lambda}$ axis direction, and the $\bar\Lambda$ in the opposite direction. The $y_{\Lambda}$ axis is perpendicular to the plane of $\Lambda$ and $e^{-}$, and the $x_{\Lambda}$ axis is defined by a right-handed coordinate system. 
The $\Lambda$ decay product, the proton, is measured in this coordinate system.}
\label{fig:helicity_frame}
\end{figure}

The exclusive process $e^+e^-\to\gamma^*/\Psi\to\Lambda\bar\Lambda\to p\bar{p}\pi^{+}\pi^{-}$ can be fully described by the $\Lambda$ scattering angle in the {center-of-mass} (CM) system of the $e^+e^-$ reaction, $\theta_{\Lambda}$, and the $p~(\bar{p})$ direction in the rest frame of its parent particle, $\boldsymbol{\hat{n}_{1}}$($\boldsymbol{\hat{n}_{2}}$). 
Here $\gamma^*/\Psi$ represents that the process $e^+e^- \to\Lambda\bar\Lambda$ is produced by pure EM process or $\psi$ resonance.
{The components of these vectors are expressed using a coordinate system ($x_\Lambda,y_\Lambda, z_\Lambda$) with the orientation shown in Fig.~\ref{fig:helicity_frame}.
 A right-handed system  for each hyperon decay is defined here, with the $z$ axis along the $\Lambda$ momentum $\textbf{p}_{\Lambda} = - \textbf{p}_{\bar{\Lambda}} = \textbf{p}$ in the CM system.} The $y$ axis is taken as the normal to the scattering plane, $\textbf{k}_{e^{-}} \times \textbf{p}_{\Lambda}$, where $\textbf{k}_{e^{-}} = - \textbf{k}_{e^{+}} = \textbf{k}$ is the electron beam momentum in the CM system. 
For the determination of the {modulus of the EM-\textit{psionic} form factors} $R^{\Psi}$~\cite{BESIII:2021cvv} and relative phase $\Delta\Phi^{\Psi}$, the angular distribution parameter $\eta$ (but not its absolute normalization) is of interest. In Ref.~\cite{Faldt:2017kgy}, the joint decay angular distribution of the process {$e^+e^-\to\gamma^*/\Psi\to\Lambda\bar\Lambda\to p\bar{p}\pi^{+}\pi^{-}$} is expressed in terms of the phase $\Delta\Phi^{\Psi}$ and the angular distribution
\begin{align}
 \label{eq:tangles:abcd}
{\cal{W}}(\boldsymbol{\xi}; \boldsymbol{\Omega}) = & 1 + \eta\cos^{2}\theta_{\Lambda} \nonumber\\
+ &\alpha_{\Lambda}\alpha_{\bar\Lambda}[\sin^{2}\theta_{\Lambda}(n_{1, x}n_{2, x} -\eta n_{1, y}n_{2, y}) \nonumber \\
+ &(\cos^{2}\theta_{\Lambda} +\eta)n_{1, z}n_{2, z}] \\
+&\alpha_{\Lambda}\alpha_{\bar\Lambda}[\sqrt{1-\eta^{2}}\cos(\Delta\Phi^{\Psi})\sin\theta_{\Lambda}\cos\theta_{\Lambda}
(n_{1, x}n_{2, z} \nonumber \\
+ & n_{1, z}n_{2, x})]\nonumber \\
+ &\sqrt{1-\eta^{2}}\sin(\Delta\Phi^{\Psi})\sin\theta_{\Lambda}\cos\theta_{\Lambda}(\alpha_{\Lambda}n_{1, y} +\alpha_{\bar\Lambda}n_{2, y}), \nonumber
\end{align}
where 
$\boldsymbol{\Omega}$ 
$= (\eta, \Delta\Phi, \alpha_\Lambda, \alpha_{\bar\Lambda})$ 
represent the production and decay parameters, the kinematic variables 
$\boldsymbol{\xi}$ 
$= (\theta_{\Lambda}, \boldsymbol{\hat{n}_1}, \boldsymbol{\hat{n}_2})$ 
describe the production and subsequent decay, and
$\alpha_{\Lambda(\bar\Lambda)}$ denotes the decay asymmetry of the $\Lambda(\bar\Lambda)\to p\pi^{-}(\bar{p}\pi^{+})$ decay process. The scattering angle distribution parameter $\eta$, is related to the ratio $R^{\Psi}$ by
\begin{equation}
R^{\Psi}=\sqrt{ \frac{\tau(1-\eta)}{1+\eta}},
\end{equation}
 where $\tau = \frac{s}{4m_{\Lambda}^2}$, $m_{\Lambda}$ is the known $\Lambda$ mass~\cite{PDG2020}, and $s$ is the square of the CM energy.
 If the initial state is unpolarized, and the production process is either strong or electromagnetic and hence parity-conserving, then a non-zero polarization is only possible in the transverse direction $y$. The polarization is given by
\begin{equation}
P_y=\frac{\sqrt{1-\eta^2}\sin\theta_{\Lambda}\cos\theta_{\Lambda}}{1+\eta\cos^2\theta_{\Lambda}}\sin(\Delta\Phi^{\Psi}).
\label{eq:pol}
\end{equation}

\begin{figure}[!htbp]
 \begin{center}
     \includegraphics[width=0.9\textwidth]{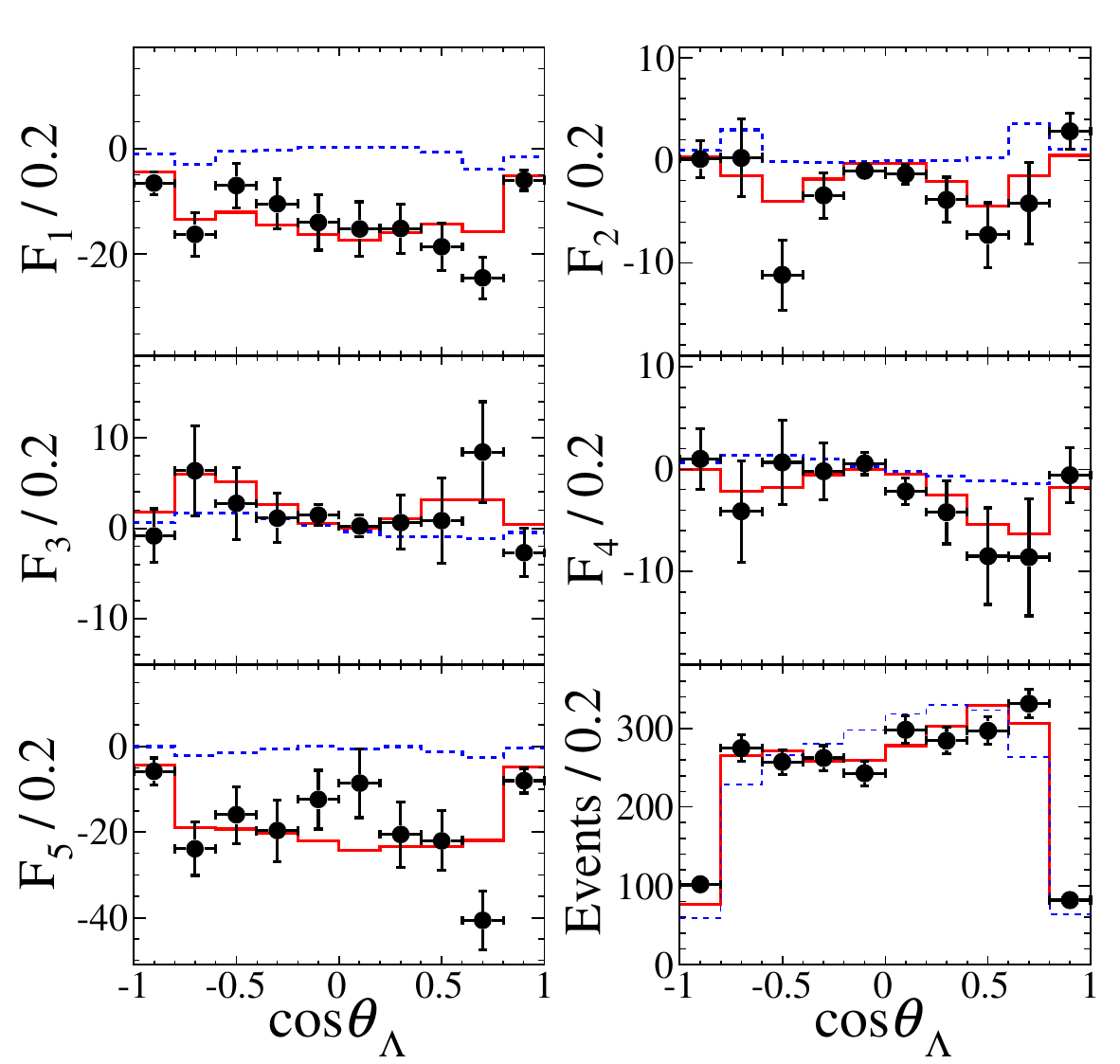}
 \end{center}
\caption{
Distributions of ${F}_{k} (k = 1, 2, ..., 5)$ moments with respect to the $\cos\theta_{\Lambda}$ and the $\cos\theta_{\Lambda}$  distribution (bottom right). The dots with error bars are data from total energy points, and the red solid lines are the weighted PHSP MC corrected by the results of global fit.
The blue dashed lines represent the distributions of the simulated events evenly distributed in phase space, without polarization.}
\label{scatter_plot::llb:projections}
\end{figure}

\begin{figure}[htbp]
\centering
\includegraphics[width=0.9\textwidth]{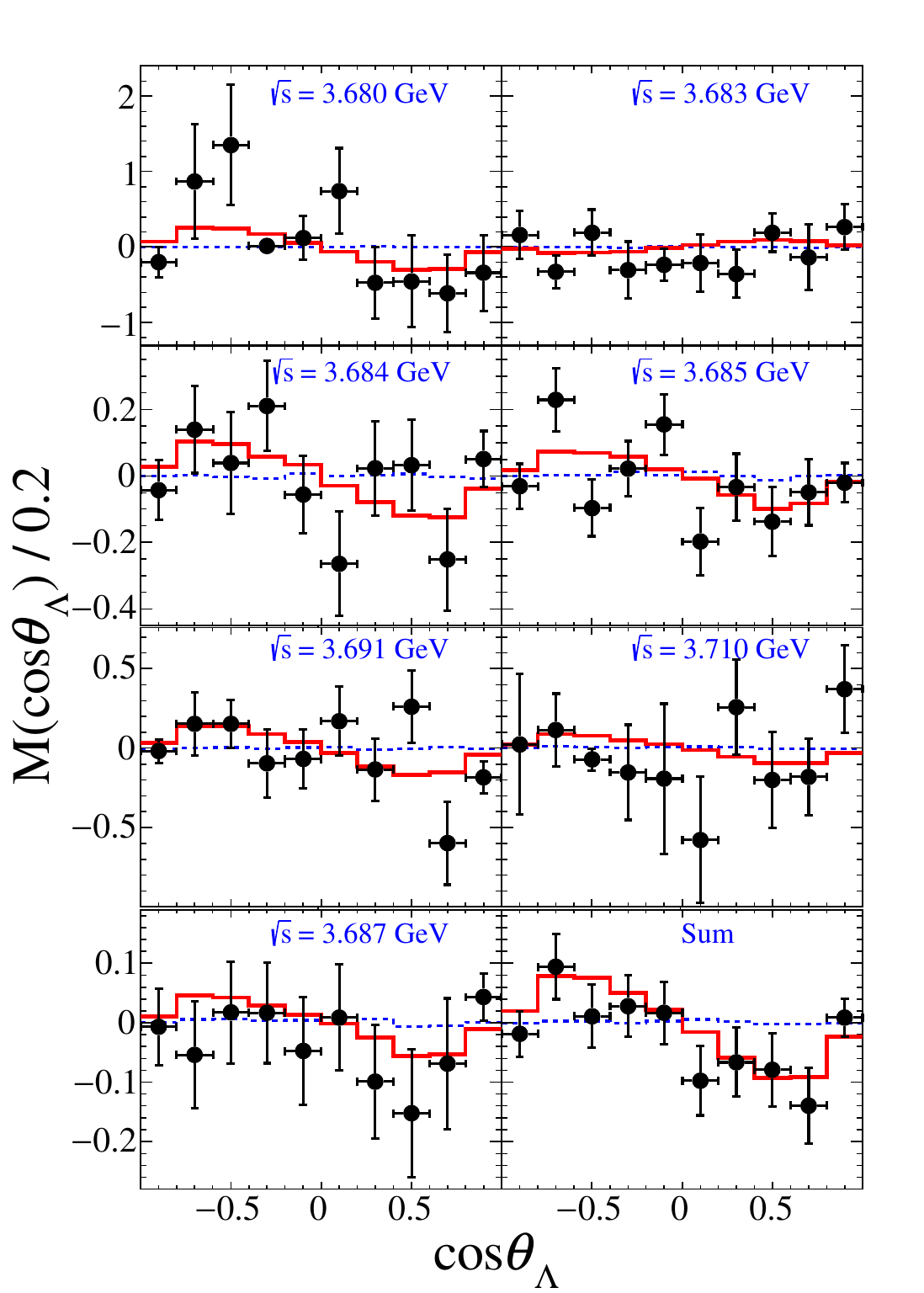}
\caption{\small {The moment $M(\cos\theta_{\Lambda})$ as a function of $\cos\theta_{\Lambda}$ for the $e^+e^-\to\Lambda\bar\Lambda$ reaction around $\sqrt{s} = 3.686$ GeV. Points with error bars are data, the red solid lines are the weighted PHSP MC corrected by the results of global fit, and the blue dashed lines represent the distributions without polarization from simulated events, evenly distributed in the phase space.}}
\label{cos}
\end{figure}

To determine the set of $\Lambda$ spin polarization parameters $\boldsymbol{\Omega}$, an unbinned maximum likelihood fit is performed to extract the decay parameters, where the decay parameters $\alpha_{\Lambda/\bar\Lambda}$ are fixed to the value 0.754 obtained from the average in Ref.~\cite{BESIII:2022yprl} assuming $CP$ conservation.
In the fit,  the likelihood function $\mathscr{L}$ is constructed from the probability function, ${\cal{P}}({\boldsymbol{\xi}}_i)$, for an event $i$ characterized by the measured angles $\boldsymbol{\xi}_i$
\begin{equation}
\mathscr{L} = \prod_{i=1}^N {\cal{P}}({\boldsymbol{\xi}}_i, {\boldsymbol{\Omega}}) = \prod_{i=1}^N {\cal{C}}{\cal{W}}({\boldsymbol{\xi}}_i, {\boldsymbol{\Omega}})\epsilon(\boldsymbol{\xi}_i), 
\end{equation}
where $N$ is the number of events in the signal region. The joint angular distribution ${\cal{W}}({\boldsymbol{\xi}}_i, {\boldsymbol{\Omega}})$ is given in Eq.~(\ref{eq:tangles:abcd}), and $\epsilon(\boldsymbol{\xi}_i)$ is the detection efficiency.
As for the ISR effect at higher energy points {3.691 and 3.710 GeV, studies based on MC simulations are performed and the contribution from ISR process is negligible, where the input cross section for $e^+e^-\to\Lambda\bar\Lambda$ for calculating the ISR effect is taken from Ref.~\cite{borncs}.} 
The normalization factor $\mathcal{C}=\frac{1}{N_\mathrm{MC}}\sum_{j=1}^{N_\mathrm{MC}} {\cal{W}}({\boldsymbol{\xi}}^{j}, {\boldsymbol{\Omega}})$ is given by the sum of the corresponding angular distribution function $\cal{W}$ using the accepted MC events $N_\mathrm{MC}$, and the difference between data and MC simulations is taken into account. 
The minimization of the function
\begin{equation}
\mathscr{S} = -\mathrm{ln}\mathscr{L}_{data} + \mathrm{ln}\mathscr{L}_{bg},
\end{equation}
is performed with the RooFit package~\cite{Roofit}.
Here, $\mathscr{L}_{data}$ is the likelihood function of events selected in the signal region, and $\mathscr{L}_{bg}$ is the likelihood function of background events determined in the sideband regions and continuum contribution, where continuum contribution is normalized by taking into account the luminosity and CM energy dependence of the cross section using the energy points at $\sqrt{s} = 3.581$ GeV, i.e.
\begin{equation}
N_{con.}=N_{3.581}\times\frac{\mathcal{L}_{Nom.}}{\mathcal{L}_{3.581}}\times\frac{s_{3.581}}{s_{Nom.}},
\label{ctmbkg}
\end{equation}
where $N_{con.}$ is the normalized event for continuum process and $N_{3.581} = 12$, $\mathcal{L}_{3.581}$ and $s_{3.581}$ are the number of observed events, the luminosity and CM energy.
The $\mathcal{L}_{Nom.}$, $s_{Nom.}$ are the luminosity and CM energy for each energy point and combined one in this work.
Note that since the statistics for this analysis is limited, the background effect is taken into account in the part of systematic uncertainty later.  
Figure~\ref{scatter_plot::llb:projections} shows distributions of the five moments \cite{Faldt:2017kgy}
\begin{align}
 \label{eq:tangles}
{{F}}_1 = &\sum_{i=1}^{N_k}(\sin^2\!\theta_{\Lambda} n^i_{1, x}n^i_{2, x} + \cos^2\!\theta_{\Lambda} n^i_{1, z}n^i_{2, z}),\nonumber\\
{{F}}_2 = &\sum_{i=1}^{N_k}\sin\!\theta_{\Lambda}\cos\!\theta_{\Lambda} (n^i_{1, x}n^i_{2, z} + n^i_{1, z}n^i_{2, x}),\nonumber\\
{{F}}_3 = &\sum_{i=1}^{N_k}\sin\!\theta_{\Lambda}\cos\!\theta_{\Lambda} n^i_{1, y},\\
{{F}}_4 = &\sum_{i=1}^{N_k}\sin\!\theta_{\Lambda}\cos\!\theta_{\Lambda} n^i_{2, y},\nonumber\\
{{F}}_5 = &\sum_{i=1}^{N_k}(n^i_{1, z}n^i_{2, z} - \sin^2\!\theta_{\Lambda} n^i_{1, y}n^i_{2, y}),\nonumber
\end{align}
with respect to $\cos\theta_{\Lambda}$, which are calculated for 10 intervals in $\cos\theta_{\Lambda}$. Here, $N_k$ is the number of events in the $k^{\rm th}$ $\cos\theta_{\Lambda}$ interval, and $i$ is the index from 1 to $N_k$.
The numerical fit results, with asymmetric uncertainties, are summarized in Table~\ref{table:sum_decay} and Table~\ref{summary}, which are consistent with theoretical predictions~\cite{PRD_116016}. 
The results presented in Table~\ref{table:sum_decay} are the combined values for  merging all scan energy points for $\psi$ resonance compared with the results from $J/\psi$ and $\psi(3770)$, which could provide more insights into the underlying production mechanism of hyperon anti-hyperon pairs at different charmonium states and different energy points.
Figure~\ref{cos} shows the result of the hyperon transverse polarization $P_{y}$ distribution,
which is consistent with the behavior of Eq.~(\ref{eq:pol}) as compared to the data. The moment given by
\begin{equation}\label{moment}
M(\cos\theta_{\Lambda}) = \frac{m}{N}\sum_{i=1}^{N_k}(n^{i}_{1,y} - n^{i}_{2,y}),
\end{equation}
is related to the polarization, and calculated for $m = 10$
intervals in $\cos\theta_{\Lambda}$. Here, $N$ represents the total number of events in the data sample. Assuming $CP$ conservation, we have $\alpha_{\Lambda} = - \alpha_{\bar\Lambda}$, 
{and the expected angular dependence is $M(\cos\theta_{\Lambda})$, which is proportional to$\sqrt{1-\eta^{2}}\alpha_{\Lambda}\!\sin\Delta\Phi^{\Psi}\cos\theta_{\Lambda}\sin\theta_{\Lambda}$ as shown in Fig.~\ref{cos} according to Eq.~(\ref{eq:tangles:abcd}). }

The relative phase of the EM-$psionic$ form factors is different from zero with a significance of 2.6$\sigma$ including systematic uncertainty,
and is estimated by comparing the likelihoods of the baseline fit and the one defined assuming no polarization. 
The effects of systematic uncertainty are estimated conservatively by varying the decay parameters $\alpha_{\Lambda} (\alpha_{\bar\Lambda})$ by one standard deviation and by considering the employed requirements, respectively, and the combination with the smallest significance is adopted.

\section{Systematic uncertainties}

\subsection{$\Lambda$ reconstruction}
The uncertainty due to the $\Lambda$ reconstruction combined with the tracking and particle identification is determined from a control sample of $J/\psi\to\Lambda\bar\Lambda$ events with the same method as used in Ref.~\cite{BESIII:2021cvv}. 

\subsection{Mass window}
The uncertainty due to the requirements on the $p\pi^-$ mass window is estimated with the smearing method as introduced in Ref.~\cite{yanliang}.

\subsection{4C kinematic fit}
The uncertainty due to 4C kinematic fit is estimated using the helix correction method mentioned in Ref.~\cite{PRD_012002}. 
We repeat the fit procedure using the MC sample corrected by the track parameters, and take the difference between both results as the related systematic uncertainty.

\subsection{Background}
The systematic uncertainty due to the background is estimated in the fit of the parameter extraction with and without the contributions of sideband and continuum backgrounds. The difference is taken as the systematic uncertainty.

\subsection{Fit method}
To validate the reliability of the fit results, an input and output check based on 400 pseudoexperiments is performed with the helicity amplitude formula {given} in Ref.~\cite{Faldt:2017kgy}. The polarization parameters measured in this analysis are used as input in the formula, and the number of events in each generated MC sample is nearly equal to the number in data sample. The differences between the input value and output value are taken as the systematic uncertainty by fitting method.

\subsection{Decay parameter}
The uncertainties from the decay parameters of $\Lambda\to p\pi^{-}$, $\alpha_{\Lambda/\bar\Lambda}$, are estimated by varying the baseline value, obtained from averaging the results in Ref.~\cite{BESIII:2022yprl}, by $\pm$1$\sigma$. The largest difference in the result is taken as the systematic uncertainty.

\subsection{Summary of systematic uncertainty}
Assuming all sources are independent,
the total systematic uncertainties on the measurement of 
the polarization parameters for the $e^+e^-\to\Lambda\bar\Lambda\to p\bar{p}\pi^{+}\pi^{-}$ process are determined by the square root of the quadratic sum of these sources as listed in Table~\ref{uncertainty}.
\begin{table}[!htbp]
\caption{\small Summary of absolute value of the systematic uncertainty of polarization parameters.}
\centering
\begin{tabular}{l c c}\hline \hline
Source 	& $\eta$   &$\Delta\Phi^{\Psi}$ ($^\circ$)\\	
\hline
$\Lambda$ reconstruction   &0.002     &0.057\\
Mass window          &0.012    &0.286      \\
4C kinematic fit     &0.001	   &0.286	 \\
Sideband Background           &0.009    &1.375    \\
Continuum Background           &0.001    &0.516    \\
Fit method		     &0.001    &0.115	 \\
Decay parameter      &0.005    &0.573    \\\hline
Total                &0.016    &1.633\\
\hline
\hline
\end{tabular}
\label{uncertainty}
\end{table}

\section{Summary}
In summary, we measure the $\Lambda$ hyperon transverse polarization in the $e^+e^-\to$ $\Lambda\bar{\Lambda}$ reaction
at CM energies between 3.68 and 3.71 GeV, using a data sample corresponding to an integrated luminosity of 333 pb$^{-1}$ collected with the BESIII detector at BEPCII. By combining the seven energy points, the relative phase and the modulus of the ratio of the EM-\textit{psionic} form factors and the angular distribution parameter are determined, 
 respectively.
 The relative phase is determined to be different from zero with a significance of 2.6$\sigma$ including the systematic uncertainty. 
 The comparison between our result and previous measurements in 
$e^+e^-\to$ $\Lambda\bar{\Lambda}$ at 2.396 GeV~\cite{BESIII:2019nep} and $J/\psi \to$ $\Lambda\bar{\Lambda}$~\cite{BESIII:2022yprl} is summarized in Table~\ref{table:sum_decay}. 
The value from the phase obtained in $\psi(3686)$ resonance is roughly consistent with the measurement from $\psi(3770)$ decay and $e^+e^-$ annihilation at $\sqrt{s} = 2.396$ GeV within the uncertainty of $1\sigma$, but deviate from the measurement with $J/\psi$ decay by 2.4$\sigma$.
The $\eta$ values are obviously different from the ones for $J/\psi$ peak and $\sqrt{s} = 2.396$ GeV, but roughly consistent with the $\psi(3770)$ one. This implies the presence of different $e^+e^-\to\Lambda\bar\Lambda$ production mechanisms.
More data samples at different energy points are needed
for a detailed understanding for the underlying $\Lambda\bar\Lambda$ production mechanism and the structure of the $\Lambda$ hyperon.

\begin{table*}[htbp]
\scriptsize
\caption{\label{table:sum_decay} The measured parameters from $\psi(3686)$ resonance for merging all scan energy points compared with previous measurements by combining the seven energy points. For each measurement, the first uncertainty is statistical and the second one is systematic.
}
\begin{tabular}{lccc} 
\hline
\hline
Para.   &$\eta$ &$\Delta\Phi^{\Psi}$ ($^\circ$) &$R^{\Psi}$  \\ \hline

This work (Sum)   &0.69$^{+0.07}_{-0.07}$ $\pm$ 0.02 &23$^{+8.8}_{-8.0}$ $\pm$ 1.6  &0.71$^{+0.10}_{-0.10}$ $\pm$ 0.03 \\
$\psi(3770)\to\Lambda\bar\Lambda$~\cite{BESIII:2021cvv} &$0.85^{+0.12}_{-0.20}$ $\pm$ 0.02  &$71^{+66}_{-46}$  $\pm$ 5  &$0.48^{+0.21}_{-0.35}$ $\pm$ 0.04 \\
$J/\psi\to\Lambda\bar\Lambda$~\cite{BESIII:2022yprl}    &0.4748 $\pm$ 0.0022 $\pm$ 0.0031 &43.09 $\pm$ 0.24 $\pm$ 0.38  &$0.8283 \pm  0.0024 \pm 0.0033$ \\
$e^{+}e^{-}\to\Lambda\bar\Lambda$ ($\sqrt{s}$ = 2.396 GeV)~\cite{BESIII:2019nep} &0.12 $\pm$ 0.14 $\pm$ 0.02  &37 $\pm$ 12 $\pm$ 6  &0.96 $\pm$ 0.14 $\pm$ 0.02
\\ 
\hline
\hline
\end{tabular}
\end{table*}
\begin{table}[!htbp]
\scriptsize
\caption{The number of observed events $N_{obs}$ (from the S region defined in Fig.~\ref{scatter_plot::llb}) and measured parameters $\eta$, $\Delta\Phi$ and $R^{\Psi}$ for each energy point. For each measurement, the first uncertainty is statistical and the second one is systematic.}
\centering
\begin{tabular}{cr@{}lcr@{}lc}
\hline
\hline
$\sqrt{s}$ (MeV)	&\multicolumn{2}{c}{$N_{obs}$}				&$\eta$  &\multicolumn{2}{c}{$\Delta\Phi$($^\circ$)}    & $R^{\Psi}$\\
\hline
	 3680    & 31&$^{+7}_{-6}$        &0.12$^{+0.46}_{-0.43}$ $\pm$ 0.02  &110.6&$^{+34.4}_{-37.8}$ $\pm$ 1.6   & 1.46$^{+0.68}_{-0.64}$ $\pm$ 0.03 \\
        3683    & 78&$^{+10}_{-9}$         &0.95$^{+0.18}_{-0.20}$ $\pm$ 0.02  &-108.9&$^{+181.6}_{-185.6}$ $\pm$ 1.6   & 0.26$^{+0.49}_{-0.54}$ $\pm$ 0.05\\
        3684     & 385&$^{+21}_{-20}$        &0.59$^{+0.18}_{-0.18}$ $\pm$ 0.02  &28.0&$^{+24.1}_{-18.3}$ $\pm$ 1.6   & 0.84$^{+0.23}_{-0.23}$ $\pm$ 0.02\\
        3685     & 831&$^{+30}_{-29}$        &0.64$^{+0.11}_{-0.12}$ $\pm$ 0.02  &20.1&$^{+12.6}_{-12.0}$ $\pm$ 1.6   & 0.77$^{+0.14}_{-0.15}$ $\pm$ 0.03\\
        3687    & 876&$^{+31}_{-30}$         &0.71$^{+0.12}_{-0.13}$ $\pm$ 0.02  &12.0&$^{+14.9}_{-13.2}$ $\pm$ 1.6   & 0.68$^{+0.16}_{-0.18}$ $\pm$ 0.03\\
        3691    & 176&$^{+14}_{-13}$         &0.83$^{+0.14}_{-0.23}$ $\pm$ 0.02  &112.9&$^{+44.1}_{-63.0}$ $\pm$ 1.6   & 0.50$^{+0.23}_{-0.37}$ $\pm$ 0.03\\
        3710     & 53&$^{+8}_{-7}$        &0.52$^{+0.38}_{-0.39}$ $\pm$ 0.02  &0.0&$^{+64.7}_{-56.7}$ $\pm$ 1.6	 & 0.93$^{+0.49}_{-0.50}$ $\pm$ 0.03\\

\hline
\hline
\end{tabular}
\label{summary}
\end{table}

\acknowledgments
The BESIII collaboration thanks the staff of BEPCII and the IHEP computing center for their strong support. This work is supported in part by National Key R\&D Program of China under Contracts Nos. 2020YFA0406400, 2020YFA0406300; National Natural Science Foundation of China (NSFC) under Contracts Nos. 
12075107, 12247101,
11635010, 11735014, 11835012, 11935015, 11935016, 11935018, 11961141012, 12022510, 12025502, 12035009, 12035013, 12061131003, 12192260, 12192261, 12192262, 12192263, 12192264, 12192265; 
the Chinese Academy of Sciences (CAS) Large-Scale Scientific Facility Program; 
Supported by the 111 Project under Grant No. B20063;
the CAS Center for Excellence in Particle Physics (CCEPP); Joint Large-Scale Scientific Facility Funds of the NSFC and CAS under Contract No. U1832207; CAS Key Research Program of Frontier Sciences under Contracts Nos. QYZDJ-SSW-SLH003, QYZDJ-SSW-SLH040; 100 Talents Program of CAS; The Institute of Nuclear and Particle Physics (INPAC) and Shanghai Key Laboratory for Particle Physics and Cosmology; ERC under Contract No. 758462; European Union's Horizon 2020 research and innovation programme under Marie Sklodowska-Curie grant agreement under Contract No. 894790; German Research Foundation DFG under Contracts Nos. 443159800, 455635585, Collaborative Research Center CRC 1044, FOR5327, GRK 2149; Istituto Nazionale di Fisica Nucleare, Italy; Ministry of Development of Turkey under Contract No. DPT2006K-120470; National Science and Technology fund; National Science Research and Innovation Fund (NSRF) via the Program Management Unit for Human Resources \& Institutional Development, Research and Innovation under Contract No. B16F640076; Olle Engkvist Foundation under Contract No. 200-0605; Polish National Science Centre under Contract No. 2019/35/O/ST2/02907; STFC (United Kingdom); Suranaree University of Technology (SUT), Thailand Science Research and Innovation (TSRI), and National Science Research and Innovation Fund (NSRF) under Contract No. 160355; The Royal Society, UK under Contracts Nos. DH140054, DH160214; The Swedish Research Council; U. S. Department of Energy under Contract No. DE-FG02-05ER41374.


\end{document}